\documentclass[showpacs,preprintnumbers,amsmath,amssymb]{revtex4}
\usepackage{}
\usepackage{graphicx}
\usepackage{amsfonts}
\usepackage{amssymb}%

\providecommand{\Journal}[4] {#1 {\bf#2}, #4 (#3)}
\providecommand{\PLB}{Phys. Lett. B} %
\providecommand{\PRL}{Phys. Rev. Lett.} %
\providecommand{\PRD}{Phys. Rev. D} \providecommand{\NT}{Nature}
\providecommand{\SCI}{Science}
\providecommand{\JC}{JCAP} %
\providecommand{\ApP}{Astropart. Phys.}%
\providecommand{\AJ}{Astrophys.J.}%
\providecommand{\Aa}{Astron. Astrophys.}%
\providecommand{\Jh} {JHEP}%
\providecommand{\IB}{ibid}%
\providecommand{\IJMP}{Int. J. Mod. Phys.}%
\providecommand{\IJMPa}{Int. J. Mod. Phys.A}%

\newcommand{\be}{\begin{equation}}
\newcommand{\ee}{\end{equation}}
\newcommand{\bea}{\begin{eqnarray}}
\newcommand{\eea}{\end{eqnarray}}

\newcommand{\hf}{\frac{1}{2}}
\newcommand{\al}{\alpha}
\newcommand{\e}{\epsilon}
\newcommand{\nn}{\nonumber\\}

\newcommand{\pt}{\partial}
\def\syjm#1#2{\phantom{}_{#1}Y_{#2}}

\begin{document}

\title{Constraints on Lorentz invariance violation from gamma-ray burst GRB090510}

\author{Zhi Xiao}
\author{Bo-Qiang Ma\footnote{Corresponding author. Email address: \texttt{mabq@phy.pku.edu.cn}}}
\affiliation{School of Physics and State Key Laboratory of Nuclear
Physics and Technology, Peking University, Beijing 100871, China}

\begin{abstract}
We obtain modified dispersion relations by requiring the vanishing
of determinant of inverse of modified photon propagators in Lorentz
invariance violation (LIV) theory. Inspired by these dispersion
relations, we give a more general dispersion relation with less
assumption and apply it to the recent observed gamma-ray burst
GRB090510 to extract various constraints on LIV parameters. We find
that the constraint on quantum gravity mass is slightly larger than
the Planck mass but is consistent with other recent observations, so
the corresponding LIV coefficient $\xi_1$ has reached the natural
order ($\mathcal {O}(1)$) as one expects. From our analysis, the linear LIV
corrections to photon group velocity might be not excluded yet.

\keywords{Lorentz invariance violation; modified dispersion
relation, gamma-ray burst}
\end{abstract}

\pacs{11.30.Cp, 11.30.Er, 11.55.Fv, 98.70.Rz} \maketitle

\maketitle
\section{Introduction\label{sec:1}}
Lorentz invariance violation (LIV or LV) has been intensively
investigated both theoretically and experimentally in recent years.
The revival passion of relativity violation in theoretical
construction originates from the attempt to compromise general
relativity with quantum mechanics. On the other hand, the
experimental searches may provide us with concrete evidence to sift
a most hopeful candidate of quantum gravity from a vast number of
theories.

From theoretical aspect, some theories expect LIV to happen at high
energies. For example, spontaneous Lorentz symmetry breaking may
happen in string theory as the perturbative string vacua is
unstable, thus some tensor fields generate nonzero vacuum
expectation values~\cite{string}. The breaking of Lorentz symmetry
also happens in other frameworks, such as loop gravity~\cite{loop},
foamy structure of spacetime~\cite{foam}, torsion in general
gravity~\cite{Yan}, etc.. More recently, Ho$\breve{\mathrm{r}}$ava
proposed a power counting renormalizable theory of
gravity~\cite{Horava} with a ``dynamical critical exponent" $z$ to
characterize the anisotropic scaling properties between space and
time. While Lorentz symmetry is breaking at high energies, it
restores when this dynamical critical exponent flows to $z=1$ at low
energies. There are also some other proposals, such as the so called
double special relativity~\cite{DSR}, which preserves relativity
principle with a nonlinear realization of Lorentz group, thus
conventional Lorentz symmetry is also broken. One striking
consequence of LIV is that the photon propagation speed is no longer
a unique constant, generally, it depends on energy and propagation
direction.

These theoretical investigations have promoted various experiments
to search for the deviation from conventional linear dispersion
relation for photons~\cite{Exp}. However, as the possible violation
effects for photons must be very tiny, the detection of these
effects present a significant challenge to experimentalists. In
addition to improve the precision of measurements to find any
possible evidence of LIV, we should also take efforts on searching
for certain accumulating processes to amplify these tiny effects.
Such idea has already been proposed on the observation of certain
astronomical objects such as gamma-ray bursts (GRB)~\cite{foam,TOF}, pulsars~\cite{pulsar} and active galactic
nuclei~(AGN)~\cite{AGN}, etc.,
and the tiny LIV effect could manifest itself through the
observation of rotation of linear polarization
(birefringence)~\cite{SME} or time of flight lag~\cite{TOF} for
photons with different energies.

The paper is organized as follows. In Section \textrm{2}, we review
certain modified photon dispersion relations derived from several
LIV models, including standard model extension (SME) with power
counting renormalizable operators~\cite{SME}, effective field theory
with dimension 5 operators~\cite{EFT} and
Ho$\breve{\mathrm{r}}$ava's anisotropic U(1) theory. In Section
\textrm{3}, we focus on time of flight analysis of GRB and try to
extract some LIV parameters from the recent observation of
GRB090510~\cite{090510}. We briefly discuss the time of flight
analysis of photons from cosmological distant objects, then we give
a general dispersion relation used conventionally in the
astrophysical analysis of LIV~\cite{AGN}. This general dispersion
relation contains those terms derived from the models in Section
\textrm{2} as special cases. We then extract constraints to linear
LIV parameters from GRB090510 to $\mathcal {O}(0.1)$, improved by 1 or 2 order
of magnitude than those in~\cite{FAT} and ~\cite{AGN,MAGIC}. From
the analysis of the time-lag formula we find that it is hard to
significantly improve the constraints 
from this simple and rough analysis, unless other time-lag effects
(like source effect~\cite{Shao}, which is a major uncertainty in the
time of flight analysis) can be clarified or other methods will be
used.

\section{Photon Dispersion Relations\label{sec:2}}
\subsection{ Background tensor field induced LIV\label{sec:a}}

A systematical treatment of LIV to incorporate particle standard
model with power counting renormalizable Lagrangian, called standard
model extension (SME), was proposed by Kosteleck\'y and Colladay in
Ref.~\cite{SME}, where the photon sector reads \bea\label{rp}
    &&
    \mathcal{L}_\mathrm{photon}=-\frac{1}{4}F_{\mu\nu}F^{\mu\nu}-\frac{1}{4}(k_F)_{\kappa\lambda\mu\nu}F^{\kappa\lambda}
         F^{\mu\nu}+\frac{1}{2}(k_{AF})_\kappa\epsilon^{\kappa\lambda\mu\nu}A_\lambda
         F_{\mu\nu}.
\eea From (\ref{rp}), we deduce the equation of motion below
\bea\label{eom}
    &&
  \partial^\alpha
    F_{\mu\alpha}+(k_F)_{\mu\alpha\beta\gamma}\partial^\alpha
    F^{\beta\gamma}+(k_{AF})^\alpha\epsilon_{\mu\alpha\beta\gamma}F^{\beta\gamma}=0.
\eea By expressing (\ref{eom}) in terms of 4-vector potential
$A_\mu$ and assuming that the Fourier decomposition
\bea\label{Fourier}
    &&A_\mu(x)\equiv~a_\mu(p)\exp(-i~p\cdot x)
    \eea is still reliable, we express (\ref{eom}) in the
    momentum space as
\bea
    &&M_{\mu\nu}(p)a^\nu(p)=0,
    \eea where \bea \label{Ma}
    &&M_{\mu\nu}(p)=\eta_{\mu\nu}~p^2-p_\mu~p_\nu-2(k_F)_{\mu\kappa\lambda\nu}p^\kappa~p^\lambda-2i(k_{AF})^\kappa\epsilon_{\mu\kappa\lambda\nu}p^\lambda.
    \eea
We impose gauge fixing condition and require the determinant of the
reduced matrix to vanish, then we can obtain an implicit function
$p^0(\vec{p})$, which is an eighth order-polynomial in $p^0$.
Otherwise one can verify that the determinant of $M_{\mu\nu}$
vanishes as a consequence of gauge invariance of equation
(\ref{eom}). For example, we use Lorentz gauge
    \be\label{lg}   \partial_\alpha~A^\alpha=0,
    \ee
in momentum space, i.e., \be p_\alpha~a^\alpha(p)=0. \ee So we have
the gauge fixed reduced matrix
    \bea \label{Mar}
    &&M^{\mathrm{gf}}_{\mu\nu}(p)=\eta_{\mu\nu}~p^2-2(k_F)_{\mu\kappa\lambda\nu}p^\kappa~p^\lambda-2i(k_{AF})^\kappa\e_{\mu\kappa\lambda\nu}p^\lambda.
    \eea
For our purpose, we just try to extract a simplified result by
assuming that (particle) rotational invariance still holds
regardless of the explicit violation of Lorentz symmetry, i.e., only
$(k_{AF})^0$ and $\alpha$ (a combination of
$(k_F)_{\kappa\lambda\mu\nu}$, for details, see Appendix or
\cite{ELV}) are nonzero. With this assumption, we have the following
matrix:
    \bea\label{rM} &&M^{\mathrm{red}}(p)=\nn &&
    \left(
      \begin{array}{cccc}
        p^2-\al\vec{p}^2 & \al p^0 p^1 & \al p^0 p^2  & \al p^0 p^3 \\
       \al p^0 p^1  & -(p^2+\al((p^0)^2+\vec{p}^2-(p^1)^2))& \al p^1 p^2+2ik_{AF}^0 p^3   & \al p^1 p^3-2ik_{AF}^0 p^2 \\
       \al p^0 p^2  & \al p^1 p^2-2ik_{AF}^0 p^3 & -(p^2+\al((p^0)^2+\vec{p}^2-(p^2)^2))& \al p^2 p^3+2ik_{AF}^0 p^1 \\
       \al p^0 p^3  & \al p^1 p^3+2ik_{AF}^0 p^2 &\al p^2 p^3-2ik_{AF}^0 p^1 & -(p^2+\al(p^0)^2+\vec{p}^2-(p^3)^2)) \\
      \end{array}
    \right).
    \eea
Its determinant reads
    \bea && \mathrm{det}(M^{\mathrm{red}}(p))=\left\{4(k_{AF}^0)^2\vec{p}^2-\left((1+\al
    )(p^0)^2-(1-\al)
    \vec{p}^2\right)^2\right\}(1+\al)(p^2)^2.
    \eea
By requiring $\mathrm{det}(M^{\mathrm{red}}(p))=0$ (otherwise there
would be no solution for a photon field), we have two dispersion
relations, one is the conventional $p^2=0$ and the other is
\be\label{ucdr}  (p^0)^2=\frac{1}{(1+\al)}\left((1-\al)
    \vec{p}^2\pm2k_{AF}|\vec{p}|\right).
    \ee
We can also use another equivalent method to obtain these two
dispersion relations. First, we rewrite (\ref{rp}) in an explicitly
quadratic form in the photon field, i.e.,
    \bea\label{frl} && \mathcal{L}_\mathrm{photon}=-\frac{1}{4}F_{\mu\nu}F^{\mu\nu}-\frac{1}{4}(k_F)_{\kappa\lambda\mu\nu}F^{\kappa\lambda}
         F^{\mu\nu}+\frac{1}{2}(k_{AF})_\kappa\e^{\kappa\lambda\mu\nu}A_\lambda
         F_{\mu\nu}-\frac{1}{2\xi}(\pt\cdot A)^2\nn  && ~~~~=-\hf~\pt_\mu A_\nu\left(F^{\mu\nu}+2(k_F)^{\kappa\lambda\mu\nu}\pt_\kappa A_\lambda+\frac{\eta^{\mu\nu}}{\xi}\pt\cdot A-2\e^{\kappa\lambda\mu\nu}(k_{AF})_\kappa A_\lambda\right)\nn   && ~~~~
         =\mathrm{total~~derivative}+\hf~A_\nu(\mathrm{D_F^{-1}})^{\nu\lambda}A_\lambda,
     \eea
where we have added gauge fixing terms $-\frac{1}{2\xi}(\pt\cdot
     A)^2$ in (\ref{frl}) and
     \bea\label{ivp}&&(\mathrm{D_F^{-1}})^{\nu\lambda}\equiv\left(\Box\eta^{\nu\lambda}-\pt^\nu\pt^\lambda(1-\frac{1}{\xi})-2(k_F)^{\nu\mu\kappa\lambda}\pt^\mu\pt^\kappa-2\e^{\nu\mu\kappa\lambda}(k_{AF})_\mu\pt^\kappa\right).
     \eea
Using the same Ansatz (\ref{Fourier}), we define a matrix $\Sigma$
in momentum space\bea\label{ivpp}&&
      \Sigma(p)_{\nu\rho}=-\left(p^2\eta_{\nu\rho}-(1-\frac{1}{\xi})p_\nu p_\rho\right)+2(k_F)_{\nu\mu\kappa\rho}p^\mu
      p^\kappa -2i\e_{\nu\mu\kappa\rho}p^\mu(k_{AF})^\kappa.
      \eea
From the conventional free field theory, the differential operator
inside the two fields in the quadratic form of certain Lagrangian
(e.g., (\ref{frl})) is just the inverse of free field propagator in
position space (see \cite{qft} or \cite{csf} ), thus (\ref{ivpp}) is
just the inverse of photon propagator expressed in momentum space.
We know that generally the inverse of propagator is just the
dispersion relation, from which one can find the pole of the
corresponding particle, so we expect that the determinant of
(\ref{ivpp}) in case of $\xi\rightarrow\infty$ (i.e., without gauge
fixing) is zero.  Similarly, we can find the explicit dispersion
relation in a special gauge by choosing the corresponding specific
value of $\xi$. For example, we find that
$\Sigma(p)_{\nu\rho}|_{\xi=1}=-M^{\mathrm{gf}}(p)_{\nu\rho}$
($\xi=1$ is just the Lorentz gauge (\ref{lg}) used to obtain
(\ref{rM}), and this choice can avoid the inequivalent gauge choice
comparison. As pointed out in~\cite{SME}, different gauge choices
are inequivalent with each other in the LIV electrodynamics). So in
the rotational invariant case, this matrix can also lead to
(\ref{ucdr}) and the conventional dispersion relation. We mention
here that similar method to obtain photon propagator in the SME
framework has also been obtained recently in~\cite{gp}, with a more
systematic and complete treatment.

The leading order nonrenormalizable LIV operators (dimension 5) were
systematically studied in ~\cite{EFT}, where Myers and Pospelov also
introduced explicitly a timelike four-vector $n^a$ to take LIV into
account, thus this theory can be regarded as a leading
nonrenomalizable part of SME. Since we are only interested in the
study of the consequence of LIV to the propagation of GRB, we focus
our attention only on photon field there. The corresponding
Lagrangian is \bea\label{Df}&&
       \delta\mathcal{L}_{\mathrm{photon}}=\frac{\xi}{2M_{Pl}}\e^{\mu\nu\kappa\rho}n^\al
       F_{\al\rho}n\cdot\pt(n_\kappa F_{\mu\nu}).
      \eea
We write it in another equivalent form, i.e.\bea\label{spl}&&
       \mathcal{L}_{\mathrm{photon}}=\hf~A_\nu\left(\Box\eta^{\nu\rho}-\frac{2\xi}{M_{Pl}}n\cdot\pt(n\cdot\pt~n_\kappa\pt_\mu\e^{\nu\mu\kappa\rho}+n_\kappa\pt_\mu\pt_\al\e^{\nu\mu\kappa\al}n^\rho)\right)A_\rho+\mathrm{total~~derivative},
      \eea
where we have added the Lorentz gauge fixing term. Then by
performing the same procedure as before, we have the reduced inverse
of propagator\bea\label{inv} &&
      \Pi(p)^{\nu\rho}=-p^2\eta^{\nu\rho}-\frac{2i\xi}{M_{Pl}}(\e^{\nu\mu0\rho}p_0^2p_\mu+\e^{\nu\mu0\al}p_0p_\mu p_\al\delta^\rho_0 )
      \eea
when expressing explicitly the time-like four-vector $n$ in a
preferred frames as $n^\rho=(1,0,0,0)$. Then by imposing\be
\mathrm{det}\Pi(p)=\mathrm{det}\left(
                                   \begin{array}{cccc}
                                    -p^2 & 0 & 0 & 0 \\
                                     0 & p^2 & -i\frac{2\xi}{M_{Pl}}(p^0)^2p^3 & i\frac{2\xi}{M_{Pl}}(p^0)^2p^2 \\
                                     0 & i\frac{2\xi}{M_{Pl}}(p^0)^2p^3 & p^2  & -i\frac{2\xi}{M_{Pl}}(p^0)^2p^1 \\
                                     0 & -i\frac{2\xi}{M_{Pl}}(p^0)^2p^2 &  i\frac{2\xi}{M_{Pl}}(p^0)^2p^1 & p^2 \\
                                   \end{array})
                                 \right)=p^4\left((\frac{2\xi}{M_{Pl}}\vec{p})^2(p^0)^4-p^4\right)=0,
      \ee
we obtain the dispersion relation
      \be\label{fdf}
      (p^0)^2=\vec{p}^2\pm\frac{2\xi}{M_{Pl}}(p^0)^2|\vec{p}|,\ee
which was obtained in~\cite{EFT} plus the conventional one $p^2=0$.

\subsection{ Anisotropic scaling induced LIV  \label{sec:b}}
Now we turn to another framework of LIV proposed recently by
Ho$\breve{\mathrm{r}}$ava~\cite{Horava}. His original proposal was
to provide a UV completion of quantum theory of gravity. Lorentz
symmetry appears naturally in this theory when the dynamical
critical exponent flows to $z=1$ at low energies. While at high
energies, space and time present anisotropic scaling \bea
t\rightarrow\lambda^zt, \quad \vec{r}\rightarrow\lambda\vec{r}, \eea
thus Lorentz symmetry breaks down. However, this formalism does not
break spatial isotropy, thus there is no need to assume a special
background field configuration to realize rotational invariance,
unlike the background tensor formalism discussed above. Aside from
gravity, Ho$\breve{\mathrm{r}}$ava also constructed an anisotropic
Yang-Mills theory with critical spatial dimension
$D=4$~\cite{Horava}. As Chen and Huang recently gave a general
construction of bosonic field theory demonstrating this anisotropic
scaling behavior~\cite{HYM}, we follow this new approach instead of
\cite{Horava}. In the new formalism, the photon action reads
\bea\label{anf}
 S=\hf\int~dtd^Dx\frac{1}{g_E^2}\left(\vec{E}^2-\sum_{J\geq2}\frac{1}{g_E^{J-2}}\sum^{n_J}_{n=0}(-1)^n\frac{\lambda_{J,n}}{M^{2n+\hf(D+1)(J-2)}}\pt^{2n}\star
 F^J\right).
\eea For simplicity, we consider the case $z=2$ and $D=3$. Then one
immediately reads from the action that the scaling dimensions of the
couplings are \bea&& [g_E]_s=\hf(z-D)+1 ,\quad
[\lambda_{J,n}]_s=z+D+\hf(z-D-2)J-2n.\eea Thus in this case the
renormalizable condition ($[g_E]_s\geq0:z\geq~D-2$) for $\vec{E}$ is
automatically satisfied. Actually, it is superrenormalizable. If the
critical dimension (i.e., $[g_E]_s=0$) is $D=3$, then $z$ must equal
to 1, which just corresponds to the conventional Lorentz invariant
gauge theory. Renormalizability also imposes the condition
$[\lambda_{J,n}]_s\geq0$, and for a free field theory, $J=2$,
$n\leq~z-1=1$. For simplicity, we set $\lambda_{2,0}=\hf$, then the
free Lagrangian (with gauge fixing term) is \bea&&
 \mathcal{L}_\mathrm{free}=\frac{1}{g_E^2}\left(\vec{E}^2-\hf F_{ij}F^{ij}-\frac{\lambda_{2,1}}{M^2}(\pt_iF_{ik}\cdot\pt_jF_{jk}+\pt_iF_{jk}\cdot\pt_iF_{jk})\right)-\frac{1}{\xi}(\pt\cdot A)^2\nn
 \nn&&
 =\frac{1}{2g_E^2}A_\nu\left\{\left(\Box\eta^{\nu\rho}-\pt^\nu\pt^\rho(1-\frac{1}{\xi})\right)-\frac{3\lambda_{2,1}}{M^2}\Delta(\Delta\delta_{kj}-\pt_k\pt_j)\delta^\nu_k\delta^\rho_j\right\}A_\rho.
 \eea
By performing the same trick, we can obtain \be
\Gamma(p)^{\nu\rho}=-p^2\eta^{\nu\rho}+\frac{3\lambda_{2,1}}{M^2}\delta^\nu_k\delta^\rho_j(p_k
p_j-\vec{p}^2\delta_{jk}), \ee and the corresponding dispersion
relations read \bea\label{Hdr} p^2=0,\quad
(p^0)^2=\vec{p}^2(1+\frac{3\lambda_{2,1}}{M^2}\vec{p}^2). \eea

\section{Time of flight Analysis of GRB in LIV theory}
In this section, we discuss the LIV effect on the observed GRBs,
focusing especially on the time of flight of $\gamma$ rays.
Gamma-ray bursts (GRBs) are sudden, intense flashes originating from
distant galaxies with cosmological distances, and they are the most
luminous electromagnetic events we ever known. As already shown in
the above formulas, LIV can modify conventional Maxwell equations
and hence leads to modified dispersion relation in addition to the
conventional one. However, due to the large mass scale suppression,
these LIV effects must be very tiny to account for the conventional
stringent terrestrial test. Fortunately, as first pointed out
in~\cite{foam}, the cosmological origin plus high energy and the
millisecond time structure of GRB make GRB an ideal object to
observe the possible minuscule effects of LIV. Actually, many known
stringent constraints to LIV parameters were drawn from astronomical
observations, such as AGN~\cite{AGN}, ultrahigh-energy cosmic rays
(UHECR)~\cite{CR}, CMB~\cite{cmb,UT}, etc.. The LIV induced modified
dispersion relation can lead to many interesting phenomena. The most
apparent consequence is the frequency dependence of photon group
velocity, though this is not always the case. For example, if only
$k_F\neq0$ in the SME framework, photons propagate independently
with their energies. So if photons with different energies are
emitted simultaneously, this frequency dispersion of group velocity
then leads to the so called time-lag phenomenon. In addition to time
lag, certain models, e.g. SME, indicate that photons with
independent polarizations obey distinct dispersion relations. This
was demonstrated in the above two models, see (\ref{ucdr}) and
(\ref{fdf}). All these models involve a conventional mode with an
extraordinary helicity dependent one, thus can lead to the so called
vacuum birefringence effects~\cite{SME,CFJ}. The tiny changes in
polarization grow linearly with propagation distance and hence can
be accumulated to be observable for cosmological sources. This can
provide a sensitive probe to LIV~\cite{cmb,UT,Bire,SME,CFJ}. Aside
from purely kinematic effects, the tiny LIV correction to dispersion
relation can also dramatically change the thresholds of high-energy
particle reactions, hence leads to distinct particle spectrum of
UHECR with respect to that of Lorentz invariance cases. The
observation of this spectrum can provide a unique signature of
LIV~\cite{insp}. On the contrary, the nonobservation of these
effects can put very stringent constraints to LIV
parameters~\cite{CR}. Below we will primarily discuss the GRB
time-lag caused by LIV.

First we make a brief review of the formula used in the description
of GRB photon time-lag with respect to the source redshift. For an
isotropic and homogeneous universe, one can derive a differential
relation \bea\label{dred}
dt=-\frac{dz}{H_0(1+z)\sqrt{\Omega_\Lambda+\Omega_K(1+z)^2+\Omega_M(1+z)^3+\Omega_R(1+z)^4}}
\eea from the Friedman equation\bea\label{Fried}
(\frac{\dot{a}}{a})^2+\frac{K}{a^2}=\frac{8\pi G_N \rho}{3}, \eea
see~\cite{SW} for details, where the present-day Hubble constant
$H_0\simeq 71~\mathrm{km/s/Mpc}$ and $\Omega_K=0$ for a nearly flat
universe. The matter density $\Omega_M\simeq 0.27$, radiation
density $\Omega_R\simeq 0$ and vacuum energy density
$\Omega_\Lambda\simeq 0.73$ are the cosmological parameters
evaluated today. We note that (\ref{dred}) is the standard result
derived from general relativity which is a locally Lorentz invariant
(LI) theory. So in dealing with photon time-lag below, we implicitly
assume that the gravity side is untouched. Though a unified
treatment should also include the change of gravity due to possible
LIV effects hence may also change (\ref{dred}) and the time-lag
formula used below.

Then we make a general assumption of photon dispersion
relation\bea\label{gdr1} E^2=f(p; \mathrm{M}, \{\xi_i\}), \eea where
$f(p; \mathrm{M}$, $\{\xi_i\})$ is a general function of $p$
($p=|\vec{p}|$), some unknown large scale $\mathrm{M}$ relevant to
LIV and a set of parameters $\{\xi_i\}$. Inspired by the dispersion
relations (\ref{fdf}) and (\ref{Hdr}) derived from particular models
discussed above and the fact that LIV corrections must be very tiny
at low energies, we assume that the expansion of (\ref{gdr1}) around
the conventional dispersion relation $E^2=p^2$ is \bea\label{gdr2}
E^2=p^2(1+\sum_{i=1}^N \xi_i(\frac{p}{\mathrm{M}})^i), \eea where
$N$ is a large number marking the precision of our expansion. Thus
(\ref{fdf}) and (\ref{Hdr}) can be regarded as just two special
cases of (\ref{gdr2}), i.e., only $\xi_1\neq0$ and $\xi_2\neq0$
respectively (where (\ref{fdf}) just adds helicity dependence
assumption of LIV). In addition to its generality, the reason for
beginning with (\ref{gdr2}) instead of those particular models is
that various experiments have already ruled out (\ref{fdf}) to a
convincing level (see~\cite{susy} and a recent review~\cite{MNC}).
Indeed, dimension 5 LIV operators have suffered very stringent
constraints both from frequency-dependent birefringence test with
GRB~\cite{Earbi,CQG}, Crab Nebula~\cite{NC}, and the UHECR spectrum
analysis under special assumptions~\cite{CR} (i.e. LIV corrections
to electron dispersion relations are smaller than those of photon
ones, e.g., in the Liouville string models of foamy structure of
space-time~\cite{LDS}, where only neutral gauge bosons receive
quantum-gravity corrections).

By taking into account of the expansion of universe~\cite{Jac} and
the assumption that gravity side remains intact,
the time-lag led by modified dispersion relation (\ref{gdr2}) with
leading order LIV correction of order $n$ is \be\label{gtl} \delta
t=\frac{1+n}{2}\xi_n\frac{\delta{
E_0^n}}{M^n}\int_0^z\frac{(1+z')^n}{h(z')}dz' ,\ee where\be
h(z)=H_0\sqrt{\Omega_\Lambda+\Omega_K(1+z)^2+\Omega_M(1+z)^3+\Omega_R(1+z)^4},\ee
$E_0$ is the redshifted photon energy observed on earth. $\delta
E_0^n=E_l^n-E_h^n$, where $E_l$ and $E_h$ denote lower and higher
energies of observed photons respectively in the time delay.
The $n$-th order correction corresponds to dimension $n+4$ LIV
operators. This can be seen from another formula
\bea\label{Kost}\delta t=\delta
w^{d-4}\int_0^z\frac{(1+z')^{d-4}}{h(z')}dz'\sum_{jm}\syjm{0}{jm}(\hat{n})k^{(d)}_{(I)jm},
 \eea which is suitable to
the analysis of time-lag in the SME framework given recently by
Kosteleck$\acute{\mathrm{y}}$ and Mewes~\cite{care}.

Before we discuss the linear ($n=1$) and the quadratic ($n=2$)
corrections to the photon dispersion relation, we first utilize the
observed time delay in GRB090510 located at redshift
$z=0.903\pm0.003$ to give a rough estimate to photon mass. As is
well known, photon mass is represented by dimension 2 operator
(quadratic in photon fields) and may spoil gauge invariance (this is
not the case in Chern-Simons theory~\cite{csf} in space-time
dimension 3). However, the presence of photon mass does not
necessarily implies LIV. In a LI theory (Proca's theory), the
existence of a unique speed $c$ could be regarded as the limiting
speed of light for arbitrary-high energy photons. So photon mass is
absent in a LIV theory (e.g. SME~\cite{Earbi}) if gauge invariance
is still valid. But we can still give a rough estimate of its
magnitude by using (\ref{Kost}) and the fact that the bulk of the
photons above 30~MeV arrived $258\pm34$~ms later than those below
1~MeV~\cite{090510} in the observation of GRB090510 :\be\label{mass}
\sum_{jm}\syjm{0}{jm}(\hat{n})k^{(2)}_{(I)jm}\leq
1.4801*10^{-24}~\mathrm{GeV^2}. \ee This could be translated to the
photon mass bound as $m_\gamma \leq 1.217*10^{-3}~\mathrm{eV}$, much
larger than the mass upper bound given in~\cite{pdg}, $m_\gamma
\leq1*10^{-18}~\mathrm{eV}$. This confirms the remarks given
in~\cite{pgm}: ``(departures of electrostatic and magnetostatic
fields from the gauge invariant one) give more sensitive ways to
detect a photon mass than the observation of velocity dispersion."
Of course, one can obtain an effective mass bound comparable to this
as $m_\gamma \leq 1.217*10^{-19}~\mathrm{eV}$~\cite{UT}, but the
origin is different. From the bound derived we see that one would
need a much larger photon mass to explain the time of flight data if
one does not introduce the LIV effect (or other effect, e.g. source
effect). More than that, as the presence of photon mass indicates
that high energy photons propagate faster than low energy ones, this
time advance of high energy photons may cancel possible time-lag
induced by certain LIV models ($\xi_i<0$ in (\ref{gdr2}), see also
(\ref{mgv})). Thus mass effects may conspire with LIV effects to
produce a nearly nonobservation of time-lag in certain time of
flight analysis~\cite{GRB}. On the contrary, in some scenario with
$\xi_i>0$, the time-lag might be caused by the combined effects of
mass and LIV, thus the situation is still complicated. Fortunately,
due to the high precision laboratory experiment~\cite{phoma}
(constrain $m_\gamma$ to $10^{-17}~\mathrm{eV}$ level), those
scenarios mentioned above do not happen and we can safely ignore the
mass effects in our discussion about LIV constraints drawn from the
time of flight analysis of GRB090510.

Without the trouble of possible mass effects, we can then securely
discuss LIV effects in the time-lag phenomena below. As a byproduct
of (\ref{Kost}), we give a rough estimate to mass dimension 3 LIV
operators\be
\sum_{jm}\syjm{0}{jm}(\hat{n})k^{(3)}_{(I)jm}\leq1.1558*10^{-21}~\mathrm{GeV}.
\ee We see that this bound is comparable to that obtained from the
LIV effects on Schumann resonances in a natural earth-ionosphere
cavity~\cite{Mews}, though it is much weaker than the other
astronomical constraints~\cite{cmb,UT} (which are constrained to
less than $10^{-43}~\mathrm{GeV}$).

Then we turn to nonrenormalizable LIV operators but using formula
(\ref{gtl}) of time-lag instead, as it is more suitable to our
simple analysis based on general dispersion relation (\ref{gdr2}).
As usual, we only discuss energy dependent corrections to photon
group velocity to the quadratic level.  Before looking into details,
from (\ref{gdr2}) we derive modified group velocity\bea\label{mgv}&&
v_g\equiv\frac{\pt E}{\pt
p}=\frac{p}{E}\left(1+\hf(i+2)~\xi_i(\frac{p}{M})^i\right)\nn &&~~~~
=\frac{\left(1+\hf(i+2)~\xi_i(\frac{p}{M})^i\right)}{\sqrt{1+\xi_i(\frac{p}{M})^i}}\nn
&&~~~~\simeq1+\hf(i+1)~\xi_i(\frac{p}{M})^i, \eea where Einstein sum
over index $i$ from 1 to N  is indicated. Then by the same procedure
in~\cite{Jac}, we can give a time-lag formula which is accurate to
2nd order of $\frac{E}{M}$ and first order in $\Delta z$ as
\bea\label{tlp} \delta
t=\xi_1\frac{E_l-E_h}{M}\int_0^z\frac{(1+z')}{h(z')}dz'+\frac{3}{8}(4\xi_2-\xi_1^2)\frac{E_l^2-E_h^2}{M^2}\int_0^z\frac{(1+z')^2}{h(z')}dz'.
\eea  It can be seen that (\ref{tlp}) is consistent with (\ref{gtl})
when $\xi_2=0$ and $\xi_1=0$ respectively for linear (though the
linear correction is calculated to second order in the large mass
suppression, it will be checked that this can not improve the linear
constraints any more, thus in practical calculation, (\ref{gtl}) is
enough) and quadratic corrections to group velocity.

For linear energy dependent correction to 1st order, we derive from
the most conservative claim that, the bulk of the photons above
30~MeV arrived $258\pm34$~ms later than those below
1~MeV~\cite{090510}, the LIV scale
$\frac{M}{\xi_1}\sim-5.02689*10^{16}~\mathrm{GeV}$, which is 3-order
less than the Planck scale if $|\xi_1|$ is of order 1. However, if
utilizing the more stringent claim that, a single highest detected
photon from GRB090510 with $31~\mathrm{GeV}$ arrives 0.179 s
later than the main LAT emission above $100~\mathrm{MeV}$, we can
deduce a significant higher quantum gravity mass
scale\bea\label{lms}
\frac{M}{\xi_1}\sim-7.72017*10^{19}~\mathrm{GeV},\eea where the
minus sign indicates the fact that photons with higher energies
propagate slower than lower ones as already mentioned in the
discussion of photon mass ($\xi_1<0$). By direct calculation of
solving 2nd order equation of $\frac{M}{\xi_1}$ (i.e. setting
$\xi_2=0$ in (\ref{tlp})), we find that this can not improve the
result any more as mentioned. This illustrates that in a rough
estimate of linear correction to group velocity, there is no need to
take into account 2nd order correction of $(\frac{M}{\xi_1})^2$ as
(\ref{tlp}). The result (\ref{lms}) means that linear correction
gives a LIV mass scale nearly 6.32 $M_{\mathrm{Pl}}$ if
$\xi_1\sim~\mathcal {O}(1)$, which is very close to that
of~\cite{090510}. Of course, if one chooses other data from the
Table 2 in~\cite{090510}, one can obtain the same conclusion that
quantum-gravity mass scale is significantly above the Planck mass
(at most of order 102 $M_{\mathrm{Pl}}$~\cite{090510}) from this
simple analysis. This conclusion is nothing more than a translation
of the claim that the constraint on the linear energy dependent LIV
parameter $|\xi_1|$ can be placed in the range $10^{-1}\sim10^{-2}$,
if we regard $M\sim~M_{\mathrm{Pl}}$. If this constraint is
confirmed by other astrophysical observations, then it puts the
constraints at least 2 order of magnitude stronger than~\cite{AGN}
and~\cite{MAGIC}, which gives $|\xi_1|<17$ and $|\xi_1|<58$
respectively. However, these constraints are not stronger enough as
those obtained in~\cite{CR} extracted from the UHECR spectrum and
those in~\cite{Earbi} from the frequency-dependent helicity
observations of GRB930131 and GRB960924. However, we note that those
most stringent constraints ($\xi_1\leq10^{-14}$) up to now rely
either on particular assumptions (see~\cite{CR}) or helicity
dependent models, e.g. (\ref{fdf}). Thus it is necessary to put the
constraints obtained from (\ref{gdr2}) on the linear LIV parameter
$\xi_1$ to the same level (still a hard task as a span of 12 orders
to be conquered) from future observations. If so, we can finally
make a more firm claim that dimension 5 LIV operators can be
excluded firmly~\cite{MNC}. Then we may reach the conclusion in the
near future that either Lorentz symmetry is exact, or at high
energies there are some other symmetries such as SUSY plus CPT to
protect our low energy theory from receiving CPT odd
corrections~\cite{susy}.

For quadratic energy dependent correction, i.e. $\xi_1=0$, we obtain
from the most conservative claim mentioned above the constraints on
quantum-gravity mass scale
$M\sim~5.84718*10^{7}~\mathrm{GeV}$ if $\xi_2\sim~\mathcal {O}(1)$. While for
the single 31~GeV event, we obtain the constraint as
$\frac{M^2}{\xi_2}\sim-5.26767*10^{21}~\mathrm{GeV}^2$, thus
$M\sim~7.25787*10^{10}~\mathrm{GeV}$. It is obvious that the
constraints obtained from quadratic correction are much weaker than
those from linear one 
as quadratic correction being suppressed more than one power of $M$.
Thus in the future we should take more efforts to the search of more
stringent constraints on quadratic LIV correction to photon group
velocity. As mentioned above, the quadratic correction is produced
by dimension 6 operators, which are the leading order
nonrenormalizable CPT even LIV operators. As we known, various current
constraints to dimension 6 operators are also much weaker than dimension 5
ones~\cite{Earbi}.

Before we close this section, we observe that our results are
similar to that obtained recently in~\cite{AGN}
and~\cite{FAT,MAGIC}. We give these results in the table below:
\begin{table}[ht]\label{TLL}
\begin{center}
\begin{tabular}{c|c|c|c|c}
\hline\hline
 Source                       & Mkn501~\cite{MAGIC}& PKS 2155 - 304~\cite{AGN}& GRB080916C~\cite{FAT} & GRB090510~\cite{090510}       \\
 \hline
redshift                      & 0.034              & 0.116                    & 4.35                  & 0.900           \\
 $\delta t$(s)                & 240                & 27                       & 16.54                 & 0.179            \\
 $E_h$ to $E_l$(GeV)          & $10^{4}$ to $250$  & 600 to 210               &13.22 to $10^{-3}$     & 31 to 0.1        \\
\hline
 $M$(GeV)                     & $6.06*10^{17}$     & $7.51*10^{17}$           & $1.55*10^{18}$        & $7.72*10^{19}$   \\
 $\frac{t_{total}}{\delta t}$ & $6.01*10^{13}$     & $1.72*10^{15}$           & $2.34*10^{16}$        & $1.29*10^{18}$   \\
\hline \hline
\end{tabular}\caption{
 Where the first three rows below the source row are the data given
from~\cite{MAGIC,AGN,FAT,090510} respectively, the last 2 rows are
the linear quantum-gravity masses and total time to time-lag ratios
calculated from (\ref{gtl}).}
\end{center}
\end{table}

We find that the rough estimates about the linear mass scale are
consistent with those given by the reference above and the order of
magnitude of the linear mass scale ranges from $10^{17}$ to
$10^{19}$. It is easily seen from Table I that one can approach this
large magnitude with the advantages both from the quotient of total
time to time-lag (which originates from the cosmological distance
and short pulse nature of GRB) and the large absolute energy
difference (range from GeV to TeV). To further constrain the linear
order quantum-gravity mass hence the coefficient $\xi_1$ in the
future, we need to amplify the large ratio of total time to time
lag, as photons with energies much higher than already observed
(TeV) can not reach us from cosmological distance due to the pair
creation interaction with infrared background photons. 
One way to amplify the large time ratio is to exclude other non-LIV
induced time-lag factors, like different response time of
detectors~\cite{onset} (which slightly increase linear
quantum-gravity mass scale obtained in~\cite{090510}), or one turns
attention to other methods like spectrum analysis~\cite{CR},
otherwise the constraints can not be improved significantly. Further
more, a statistic
analysis by taking into account of statistic error
~\cite{MAGIC} and a multi-source analysis to calculate the
correlation between distance and time-lag~\cite{TOF} will make the
results more concrete.

\section{Conclusion}
In this paper, we reviewed several modified dispersion relations
from standard model extension and Ho$\breve{\mathrm{r}}$ava theory
in the photon sector. Dispersion relations are derived consistently
from the inverse of photon free propagators, without taking quantum
corrections into account. Inspired by these dispersion relations we
give a more general one (\ref{gdr2}) to avoid some particular
assumptions (e.g. helicity dependence). Then we apply this relation
to the time of flight analysis of recently reported GRB090510. We
obtain constraints on the linear LIV energy dependent coefficients
to the level of $\xi_1\sim \mathcal {O}(.1)$, which is equivalent to the
statement that the relevant linear quantum-gravity mass scale is
$M\sim~7.72*10^{19}~\mathrm{GeV}$. Using the same method we also get
the quadratic mass scale $M_q\sim~7.26*10^{10}~\mathrm{GeV}$,
denoting a much loose constraints to dimension 6 operators. These
results are consistent with those in~\cite{MAGIC,AGN,FAT,090510}. As
a byproduct of the time-lag formula (\ref{Kost}), we point out that
one can safely ignore the photon mass effect in the discussion of
LIV effects in the time-lag analysis of GRBs due to the stringent
terrestrial constraints on photon mass, and we obtain a constraints
$\sum_{jm}\syjm{0}{jm}(\hat{n})k^{(3)}_{(I)jm}\leq1.1558*10^{-21}~\mathrm{GeV}$
to the dimension 3 operator.

From our analysis, we find that though the constraints obtained are
far from reaching those from the spectrum analysis~\cite{CR} and
those from the helicity dependent analysis~\cite{Earbi}, our
analysis relies little on extra assumptions except the expansion
(\ref{gdr2}) and the formula (\ref{gtl}). Thus to exclude the linear
order quantum-gravity correction to photon group velocity is still
too early as long as the constraints to the general linear order
correction (\ref{gdr2}) have not approached the same level as
in~\cite{MNC}. We find that the results have already reached the
precision of probing Planck mass scale or even higher, slightly
better than~\cite{MAGIC,AGN}. If one can largely clarify other time
lag uncertainties like~\cite{onset}, the constraints could be
improved more.

\section*{Acknowledgments}

We thank Hongbo Hu and Bin Chen for reminding us about
references~\cite{090510,Jac} and the helpful discussions with Bin
Chen,  Alexandre Sakharov, Lijing Shao, Zhi-bo Xu, and Shouhua Zhu.
This work is partially supported by National Natural Science
Foundation of China (No.~10721063 and No.10975003), by the Key Grant
Project of Chinese Ministry of Education (No.~305001), and by the
Research Fund for the Doctoral Program of Higher Education (China).

\section*{Appendix}
The previous referred parameter $\al$, is just one of the parameters
defined from various combinations of $(k_F)_{\kappa\lambda\mu\nu}$.
These definitions arise for convenience from the consideration of
the symmetry of this tensor. From the Lagrangian\be
\delta\mathrm{L}=-\frac{1}{4}(k_F)_{\kappa\lambda\mu\nu}F^{\kappa\lambda}
         F^{\mu\nu},
\ee we find that $(k_F)_{\kappa\lambda\mu\nu}$ is antisymmetric to
the two indices $\kappa\lambda$ and $\mu\nu$ respectively, and is
symmetric to the interchange of these two pairs of indices. As we do
not want to include a conceivable $\theta$-type term proportional to
$\hf\e_{\kappa\lambda\mu\nu}F^{\kappa\lambda}
         F^{\mu\nu}$, we require that
         $\e_{\kappa\lambda\mu\nu}(k_F)^{\kappa\lambda\mu\nu}=0$.
By requiring that $(k_F)_{\kappa\lambda\mu\nu}$ is doubletraceless
as any trace term would serve merely as a redefinition of kinematic
terms and hence a field redefinition, $(k_F)_{\kappa\lambda\mu\nu}$
has the symmetry of Riemann tensor. Then we can define the
decomposition of $(k_F)_{\kappa\lambda\mu\nu}$ in terms of its
spatial and time indices, i.e.,
 \bea && (k_{DE})^{jk}\equiv-2(k_F)^{0j0k},\quad
         (k_{HB})^{il}\equiv\hf(k_F)^{jkmn}\e^{ijk}\e^{lmn}, \nn &&
         \quad
         (k_{DB})^{jk}\equiv-(k_{HE})^{kj}\equiv\hf(k_F)^{0jmn}\e^{kmn},
         \eea with Latin indices run from 1 to 3.
Moreover, we define\bea  \al_E=\frac{1}{3}\mathrm{tr}(k_{DE}),\quad
         \al_B=\frac{1}{3}\mathrm{tr}(k_{HB}),
         \eea and double tracelessness gives
         $\mathrm{tr}(k_{HB}+k_{DE})=0$, i.e.,
         $\al\equiv\al_E=-\al_B$.
So we can extract the trace term to define \be
         (\beta_E)^{jk}=(k_{DE})^{jk}-\al\delta^{jk},\quad
         -(\beta_B)^{jk}=(k_{HB})^{jk}+\al\delta^{jk}.
         \ee By using Bianchi identity
         $(k_F)_{\kappa[\lambda\mu\nu]}=0$, we have
         $\mathrm{tr}(k_{DB})=0$. With these definitions, we can
         rewrite the Lagrangian (\ref{rp}) as
\bea&&
\mathcal{L}_\mathrm{photon}=\frac{1}{2}(\vec{E}^2-\vec{B}^2)+\hf\al(\vec{E}^2+\vec{B}^2)+\hf\left((\beta_E)^{jk}E^jE^k+
(\beta_B)^{jk}B^jB^k+(k_{DB})^{jk}E^jB^k\right)\nn
 &&~~~~~~~~~~+k^0_{AF}\vec{A}\cdot\vec{B}-\phi\vec{k}_{AF}\cdot\vec{B}+\vec{k}_{AF}\cdot(\vec{A}\times\vec{E}).
         \eea

\end{document}